\documentclass[%
 reprint,
superscriptaddress,
 amsmath,amssymb,
 aps,
pra,
]{revtex4-1}

\usepackage{bbold}
\usepackage{mathptmx}
\usepackage{subfig}
\usepackage{psfrag,graphicx}
\usepackage{dcolumn}
\usepackage{amsmath,amssymb}
\usepackage{bm}
\usepackage{color}
\usepackage{latexsym}
\usepackage{epstopdf}
\usepackage{color}
\usepackage[english]{babel}
\usepackage{latexsym}
\usepackage{psfrag,graphicx}
\usepackage{amsmath}
\usepackage{amssymb}
\usepackage{amsfonts}
\usepackage{bm}
\usepackage{natbib}
\usepackage{epstopdf}
\usepackage{appendix}
\usepackage{rotating}
\usepackage[english]{babel}
\usepackage{aeguill}
\usepackage{ulem}
\usepackage[justification=justified]{caption}

\definecolor{mygrey}{gray}{0.35}
\definecolor{myblue}{rgb}{0.2,0.2,0.8}
\definecolor{myzard}{cmyk}{0,0,0.05,0}
\definecolor{mywhite}{rgb}{1,1,1}
\definecolor{mywhite}{rgb}{1,1,1}
\definecolor{myred}{rgb}{1,0.,0.3}

\usepackage[colorlinks=true,citecolor=myblue,linkcolor=myred]{hyperref}

\def\ba{\begin{align}}
\def\enda{\end{align}}
\def\bi{\begin{itemize}}
\def\ei{\end{itemize}}

\def\be{\begin{equation}}
\def\ee{\end{equation}}
\def\bea{\begin{eqnarray}}
\def\eea{\end{eqnarray}}
\def\bse{\begin{subequations}}
\def\ese{\end{subequations}}

\newcommand{\ket}[1]{|{#1}\rangle}                       
\newcommand{\bra}[1]{\langle {#1}|}                      

\newcommand{\Ignore}[1]{ }

\begin{document}

\preprint{APS/123-QED}

\title{Cooling of Many-Body Systems via Selective Interactions}

\author{R. Grimaudo}
\address{Dipartimento di Fisica e Chimica dell'Universit\`a di Palermo, Via Archirafi, 36, I-90123 Palermo, Italy}
\address{INFN, Sezione di Catania, I-95123 Catania, Italy}

\author{L. Lamata}
\address{Department of Physical Chemistry, University of the Basque Country UPV/EHU, Apartado 644, E-48080 Bilbao, Spain}

\author{E. Solano}
\address{Department of Physical Chemistry, University of the Basque Country UPV/EHU, Apartado 644, E-48080 Bilbao, Spain}
\address{IKERBASQUE, Basque Foundation for Science, Maria Diaz de Haro 3, 48013 Bilbao, Spain}
\address{Department of Physics, Shanghai University, 200444 Shanghai, China}

\author{A. Messina}
\address{INFN, Sezione di Catania, I-95123 Catania, Italy}
\address{ Dipartimento di Matematica ed Informatica dell'Universit\`a di Palermo, Via Archirafi, 34, I-90123 Palermo, Italy}

\date{\today}

\begin{abstract}

We propose a model describing $N$ spin-1/2 systems coupled through $N$-order homogeneous interaction terms, in presence of local time-dependent magnetic fields. This model can be experimentally implemented with current technologies in trapped ions and superconducting circuits. By introducing a chain of unitary transformations, we succeed in exactly converting the quantum dynamics of this system into that of $2^{N-1}$ fictitious spin-1/2 dynamical problems. We bring to light the possibility of controlling the unitary evolution of the $N$ spins generating GHZ states under specific time-dependent scenarios. Moreover, we show that by appropriately engineering the time-dependence of the coupling parameters, one may choose a specific subspace in which the $N$-spin system dynamics takes place. This dynamical feature, which we call time-dependent selective interaction, can generate a cooling effect of all spins in the system. 
\end{abstract}

\pacs{ 75.78.-n; 75.30.Et; 75.10.Jm; 71.70.Gm; 05.40.Ca; 03.65.Aa; 03.65.Sq}

\keywords{Suggested keywords}

\maketitle

\section{Introduction}

One of the attractive aspects in the physics of trapped ions and superconducting circuits stems from their dual relationship with quantum and semi-classical spin models. On one hand, we may effectively describe the dynamics of such kind of systems in terms of the language of spin systems. On the other hand, through these highly controllable technologies \cite{Devoret}, we may reproduce and implement several types of spin interactions. Thus, trapped ions and superconducting circuits provide examples of quantum simulators of the dynamical behaviour of other quantum systems in accordance with the original seminal idea of Feynman \cite{Feynman}, mathematically reformulated in terms of digital operations some years later~\cite{Lloyd}.

A fascinating formal aspect of these quantum simulations is the mathematical occurrence of local $N$-wise spin-1/2 coupling terms in the Hamiltonian. Here $N$-wise means that the interaction among the $N$ spins may be represented as an $N$-degree homogeneous multilinear polynomial in the 3$N$ dynamical variables of all the $N$ spins. Such a kind of coupling is of course alien to physical context like nuclear, atomic, and molecular physics. However, the usefulness of such $N$-spin Hamiltonian models has been recently brought to light in the treatment and the study of fermion lattice models where many-body interactions are present \cite{Casanova}. It is possible to implement many-body interactions of higher than second-order through both trapped ions techniques \cite{Barreiro, Muller} and superconducting transmon qubit arrays \cite{MezzacapoPRL113} by exploiting collective entangling operations \cite{MolSor, Monz}.
Their physical and technological importance stems from the possibility to ease several tasks of quantum information processing. In this manner, we may drive the generic many-qubit transition $\ket{-}^{\otimes N} \rightarrow \ket{+}^{\otimes N} $ to prepare multipartite Greenberger-Horne-Zeilinger (GHZ) states with a single operation and to implement stabilizer operators \cite{Muller, Nigg} with local qubit rotations. This will allow for the implementation of topological codes \cite{Kitaev}, among other effects. Finally, the interest of studying higher order interactions may be found also in their relevance in describing better physical features and dynamical aspects of complex systems \cite{Zylberberg}.

\section{The Model and its Symmetries}

In this work, we investigate the properties of a system of $N$ distinguishable spin-1/2's subject to different magnetic fields and interacting in accordance to the following specific uniform $N$-wise interactions,
\begin{equation}\label{HNspinMessina}
H=\sum_{k=1}^{N}\hbar\omega_{k}\hat{\sigma}_{k}^{z}+\gamma_{x}\prod_{k=1}^{N}\hat{\sigma}_{k}^{x}+\gamma_{y}\prod_{k=1}^{N}\hat{\sigma}_{k}^{y}+\gamma_{z}\prod_{k=1}^{N}\hat{\sigma}_{k}^{z}.
\end{equation}
Here, uniform means that no term mixing different components of different spins, e.g. $\hat{\sigma}_1^x\hat{\sigma}_2^y\hat{\sigma}_3^z \dots \hat{\sigma}_N^x,$ is present in the Hamiltonian where only three ``diagonal'' terms appear. The coupling constants $\gamma_x$, $\gamma_y$ and $\gamma_z$ quantitatively characterize these three terms. $\hat{\sigma}^x$, $\hat{\sigma}^y$ and $\hat{\sigma}^z$ are the standard Pauli matrices and $\hbar \omega_k$ is the energy separation induced in the $k$-th spin by its relative magnetic field. We are able to exactly diagonalize this model by reducing it into a set of independent problems of single spin-1/2. It is worthwhile to note that our technique may be applied even when the Hamiltonian parameters are time dependent. This circumstance provides the key to govern the dynamics of all the spins by manipulating the time-dependent magnetic field acting upon only one out of the $N$ spins. A similar approach has already been exploited to investigate dynamical aspects of two-spin systems \cite{GMN,GMIV,GBNM}. The main result reached in this paper is a dynamically generated cooling effect of the whole $N$-spin system. This effect consists in projecting the $N$ spins in their vacuum state $\ket{-}^{\otimes N}$ as a consequence of a single measurement made on an ancilla spin (e.g. the first spin) at an adequate time.

The Hamiltonian \eqref{HNspinMessina} may be exactly diagonalized by means of a chain $\mathbb{U}$ of unitary transformations after which it may be put in the following form (see Appendix \ref{Transf Prod}).
\begin{widetext}
In the case of an odd number of spins, it reads
\begin{equation}\label{HNdispariSpinDiagonalizzata}
\tilde{H}\equiv\mathbb{U}^{\dagger}H\mathbb{U}=\hbar\Biggl[\omega_{1}+\sum_{k=2}^{N}\omega_{k}\prod_{k'=2}^{k}\hat{\sigma}_{k'}^{z}\Biggr]\hat{\sigma}_{1}^{z}+\gamma_{x}\hat{\sigma}_{1}^{x}+\Biggl[(-1)^{N-1\over 2}\gamma_{y}\prod_{k=1}^{(N-1)/2}\hat{\sigma}_{2k+1}^{z}\Biggr]\hat{\sigma}_{1}^{y}+\Biggl[\gamma_{z}\prod_{k=1}^{(N-1)/2}\hat{\sigma}_{2k+1}^{z}\Biggr]\hat{\sigma}_{1}^{z},
\end{equation}
whereas for an even number of spins, it assumes the form
\begin{equation}\label{HNPariSpinDiagonalizzata}
\tilde{H}\equiv\mathbb{U}^{\dagger}H\mathbb{U}=\hbar\Biggl[\omega_{1}+\sum_{k=2}^{N}\omega_{k}\prod_{k'=2}^{k}\hat{\sigma}_{k'}^{z}\Biggr]\hat{\sigma}_{1}^{z}+\gamma_{x}\hat{\sigma}_{1}^{x}+\Biggl[(-1)^{N \over 2}\gamma_{y}\prod_{k=1}^{N/2}\hat{\sigma}_{2k}^{z}\Biggr]\hat{\sigma}_{1}^{x}+\gamma_{z}\prod_{k=1}^{N/2}\hat{\sigma}_{2k}^{z}.
\end{equation}

\end{widetext}
The total unitary operator accomplishing this chained transformations may be written as
\begin{equation}\label{OperatoreDiagonalizzanteHNspin}
\mathbb{U}
=\dfrac{1}{2^{N-1}}\prod_{k=0}^{N-2}\left[\mathbb{1}+\hat{\sigma}_{N-(k-1)}^{z}+\hat{\sigma}_{N-k}^{x}-\hat{\sigma}_{N-(k+1)}^{z}\hat{\sigma}_{N-k}^{x}\right].
\end{equation}

We see that the only dynamical variable representing the $k$-th spin with $k \neq 1$ in $\tilde{H}$ (even and odd case), is $\hat{\sigma}^z_k$ which is constant of motion for $\tilde{H}$ even if $\dfrac{\partial}{\partial t}\tilde{H} \neq 0$.
This means that we may treat all the $\hat{\sigma}^z_k$ ($k \neq 1$) as numbers (+1 or -1) and the string of values of these constants of motion identifies one specific subspace out of the $2^{N-1}$ dynamically invariant Hilbert subspaces.
Therefore, treating $\hat{\sigma}_{k}^{z}$ ($k \neq 1$) as parameters, Eqs. \eqref{HNdispariSpinDiagonalizzata} and \eqref{HNPariSpinDiagonalizzata} give us $2^{N-1}$ effective Hamiltonians of the first spin-1/2.
Exploiting the explicit expression of $\mathbb{U}(t)$, the dynamics generated by each effective Hamiltonian is turned into the dynamics of the $N$-spin system which of course will take place in a 2x2 still invariant subspace of $H$ as given by Eq. \eqref{HNspinMessina}.

We remark that, in each of such two-dimensional subspaces, the dynamics involves a specific state of the standard basis (s.b.), $\ket{s.b.}$, and the relative flipped state, that is the one identified by $\prod_{k}\hat{\sigma}_k^x\ket{s.b.}$. Then, we have, for example, subdynamics involving the following couples of states: $\ket{+}^{\otimes(N-m)}\ket{-}^{\otimes m}$ and $\ket{-}^{\otimes(N-m)}\ket{+}^{\otimes m}$ [$\hat{\sigma}_i^z\ket{\pm}=\pm\ket{\pm}$]. This means, in particular, that the dynamically invariant subspace identified by the eigenvalues $\sigma_{k}^{z}=1$ (for all possible $k \neq 1$) involves the two states $\ket{+}^{\otimes N}$ and $\ket{-}^{\otimes N}$ of the $N$-spin system. This implies the possibility of easily generating GHZ states of the $N$-spin system through this kind of interactions between the spins, as it is well known in literature \cite{Muller,MezzacapoPRL113}. Moreover, the added value of this model lies in the application of appropriately engineered (time-dependent) magnetic fields, in order to govern the transition of the spin system between the two states, or to manipulate in time the generation of specific superposition states \cite{GMN,GdCNM,MGMN}. However, to this end, a time-dependent analysis of the problem is necessary.

Up to now, we have only considered a time-independent Hamiltonian model. Nevertheless, note that the same arguments discussed above hold for a time-dependent Hamiltonian as well. The mathematical reason is that the unitary transformation operator $\mathbb{U}$ is independent of the Hamiltonian parameters. In this way, we are able to break down the time-dependent Schr\"odinger equation for our $N$-spin system into a set of $2^{N-1}$ decoupled time-dependent Schr\"odinger equations (see Appendix \ref{SR Eq}). This implies that an exactly solvable time-dependent scenario of a spin-1/2 could be an exactly solvable scenario for our $N$-spin system dynamics restricted in one of the $2^{N-1}$ dynamically invariant Hilbert subspaces. Therefore, the knowledge of exactly solvable problems of a single spin-1/2 subjected to a time-dependent magnetic field (like that brought to light by Rabi \cite{Rabi}) becomes strategic. Thus, it is of relevance that, quite recently, new strategies to single out controllable time-dependent magnetic fields have been reported \cite{GMN, GdCNM, MGMN, Mess-Nak, Das Sarma, Bagrov, KunaNaudts}, furnishing exact analytical solutions of the related dynamical problem.

\section{Physical Applications}

\subsection{Controllable Quantum Dynamics}

Let us now consider the following specialized model
\begin{equation}
H=\hbar\omega_1 \hat{\sigma}_1^z + \gamma_x \prod_k \hat{\sigma}_k^x
\end{equation}
and the initial condition $\ket{\psi(0)}=\ket{+}^{\otimes N}$ ($\hat{\sigma}^z\ket{\pm}=\pm\ket{\pm}$).
In this instance, following our previous symmetry-based analysis, the problem is reduced to the following fictitious single-spin-1/2 problem $\tilde{H}=\hbar\omega_1 \hat{\sigma}_1^z + \gamma_x \hat{\sigma}_1^x$, regardless of the parity of $N$.
If we now suppose that $\omega_1$ varies over time in such a way to produce a perfect inversion of the fictitious spin or a balanced superposition between the states $\ket{+}_1$ and $\ket{-}_1$ \cite{GMN,GdCNM,MGMN}, it means that in the language of the $N$ spins we are producing, respectively, a perfect inversion of all the spins at the same time and a GHZ state of our $N$-spin system.
These two cases are considered in the figure below where the exact probability transition $P_+^-(t)$ of finding the $N$-spin system in the state $\ket{-}^{\otimes N}$ starting from $\ket{+}^{\otimes N}$ is reported for two different time-dependences of the magnetic field acting upon the first spin, against the dimensionless time $\tau=\gamma_x t / \hbar$.
The expressions both of the magnetic fields (Figs. 1a and 1c) and the related transition probabilities (solid lines in Figs. 1b and 1d, rspectively) are analytically derived by solving exactly the single-spin-1/2 dynamical problem \cite{GMN, Mess-Nak}.
Thus, these cases represent exactly solvable time-dependent scenarios for the dynamics of the $N$-spin system restricted to the two-dimensional subspace involving the states $\ket{+}^{\otimes N}$ and $\ket{-}^{\otimes N}$.
\begin{figure}[htp]\label{fig:MF and P}

\begin{center}
\subfloat[][]{\includegraphics[width=0.22\textwidth]{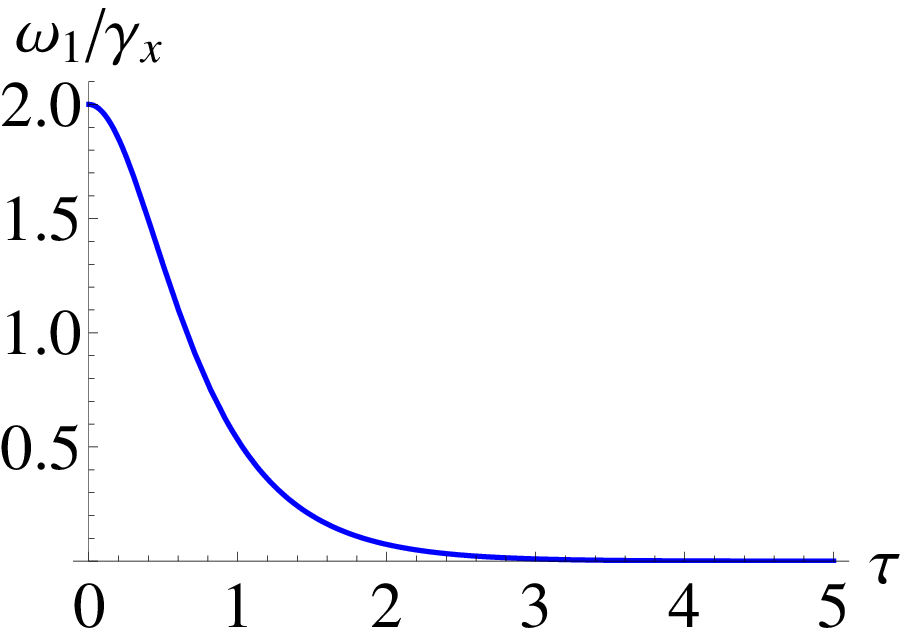}}\label{fig:Omhalf}
\quad
\subfloat[][]{\includegraphics[width=0.22\textwidth]{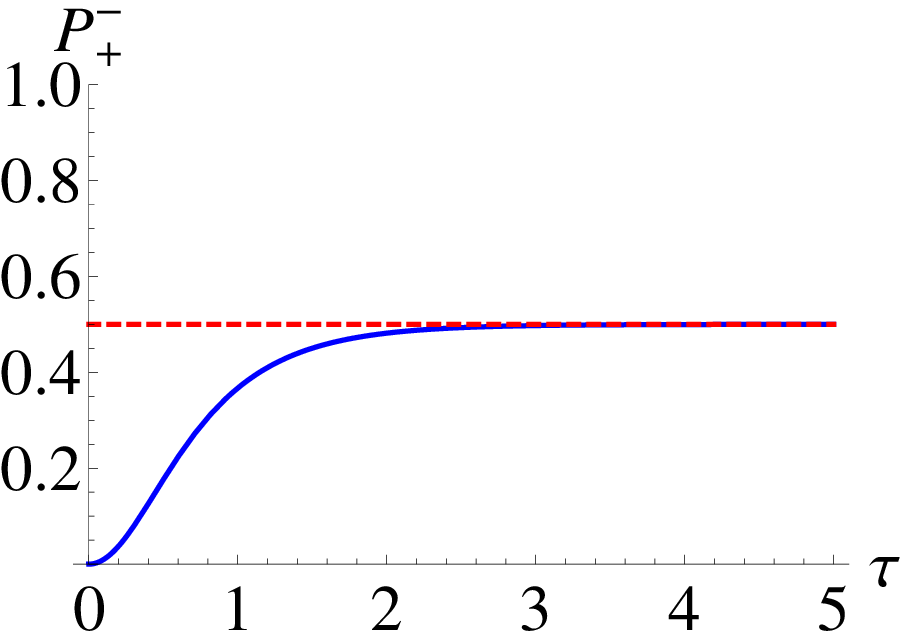}}\label{fig:Phalf}
\quad
\subfloat[][]{\includegraphics[width=0.22\textwidth]{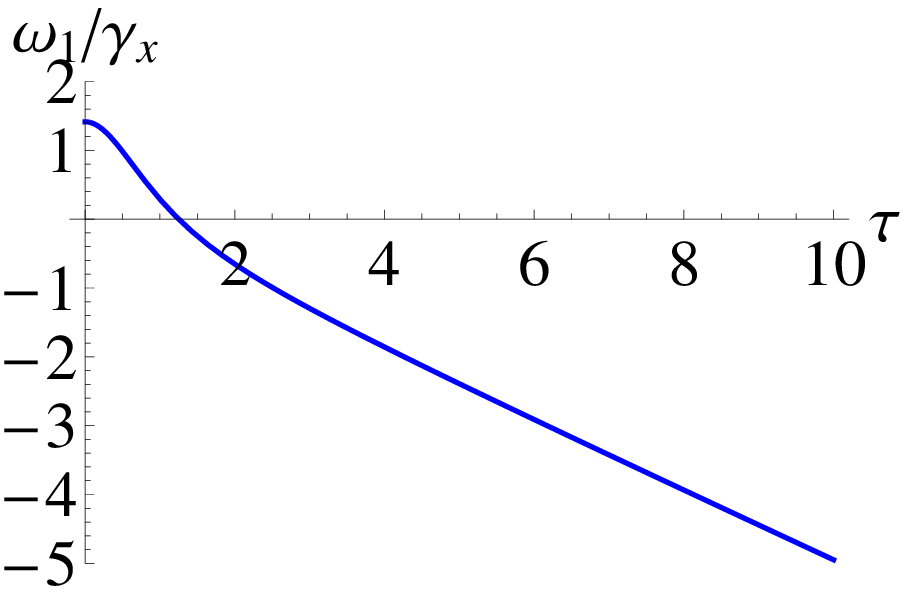}}\label{fig:Oone}
\quad
\subfloat[][]{\includegraphics[width=0.22\textwidth]{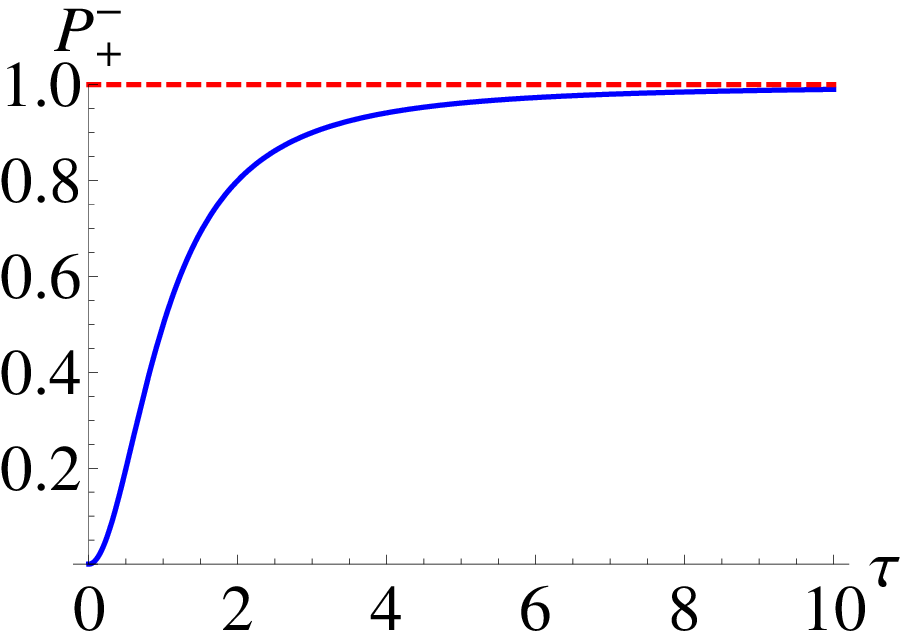}}\label{fig:Pone}
\captionsetup{justification=raggedright,format=plain,skip=4pt}%
\caption{\small(Color online) Time dependences of the magnetic field acting upon the first spin of the chain [a) and c)] with the related time behaviour of the transition probability from $\ket{+}^{\otimes N}$ to $\ket{-}^{\otimes N}$ [b) and d), respectively]. The first case consists in generating a superposition between the two states; the second case realizes an inversion of all spins in the system. The constant dashed lines in Figs. b) and d) represent $P_+^-=1/2$ and $P_+^-=1$, respectively.}
\end{center}
\end{figure}
Analogously, we may generate Rabi oscillations between the two states of the $N-$spin system involved in the subdynamics, by appropriately varying over time the parameter $\gamma_x$ \cite{Rabi,GdCNM}.
This means that, through the $N$-spin model under scrutiny, we may govern the dynamics of the whole $N$-spin system by appropriately engineering in time either the magnetic field acting upon only the first spin (ancilla qubit) or the coupling parameter or both.

\subsection{Selective interaction-based Cooling Effect}

Now we want to discuss a possible application of experimental interest aiming at attaining a \textit{cooling effect} of the whole spin system.
It is based on the possibility of selecting the invariant subspace wherein the $N$-spin-system dynamics occurs by appropriately varying over time the coupling parameter(s).
This idea was presented for the first time in \cite{SolanoSC} but developed within another physical context.
To this end, let us consider the following specialized model
\begin{equation}
H=\sum_{k=1}^N \hbar \omega_k\hat{\sigma}_k^z + \gamma_x(t) \prod_k \hat{\sigma}_k^x + \gamma_y(t) \prod_k \hat{\sigma}_k^y,
\end{equation} 
with $\gamma_x(t)=\gamma\cos(\nu t)$ and $\gamma_y(t)=\gamma\sin(\nu t)$ and for an odd number of spins.
In this way, in each subdynamics we have Rabi oscillations between the two involved standard basis states.
We know that these oscillations occur with maximum probability when the oscillation frequency of the (fictitious) transverse magnetic field ($\nu$), is equal to the characteristic frequency between the two energy levels.
Let us analyse the following conditions: 1) the characteristic frequencies of all spins are much larger than the coupling constant, $\omega_k \gg \gamma/\hbar$; 2) the oscillation frequency of the coupling constants ($\nu$) matches the resonant condition in a specific subspace.
In these instances, then, we obtain a complete oscillating behaviour in this `\textit{selected}' subspace, while in all the other ones the system dynamics is frozen since the transition probability is negligible.

To be more explicit, let us consider for simplicity three spins and the initial condition involving all the states characterized by the first spin (ancilla qubit) in the state $\ket{+}$. 
In each subspace the probability of transition from the effective state $\ket{+}$ to the effective state $\ket{-}$ reads $P_+^-(t)={(\gamma/\hbar)^2 \over (\gamma/\hbar)^2+\Delta_n^2}\sin\left({\omega_{R}\over 2}t\right)$, with $\omega_{R}=\sqrt{\Delta_n^2+(\gamma/\hbar)^2}$ and $\Delta_n=\left[\omega_{1}+\sum_{k=2}^{3}\omega_{k}\prod_{k'=2}^{k}\hat{\sigma}_{k'}^{z}\right]+\nu_n$.
Here $n$ discriminates the different sub-dynamics and $\nu_n=\pm\nu$ depending on the sub-space, as it is clear by Eq. \eqref{HNdispariSpinDiagonalizzata}.
It is easy to verify that if we assume now, for example, $\nu=-\sum_{k=1}^3\omega_k$ we have complete oscillations, that is $P_+^-(t)= \sin\left({\omega_{R}\over 2}t\right)$, only in the subspace involving the two states $\ket{+}^{\otimes 3}$ and $\ket{-}^{\otimes 3}$.
In the other subspaces, instead, providing that the condition $\omega_k \gg \gamma, \forall k$ is satisfied, the probability of transition is negligible and the dynamics is frozen, in the sense that the state remains the initial one.

This coupling-based dynamical selectivity turns out to be of particular relevance in the light of the following application of experimental interest.
In the case of three spins under scrutiny, for the sake of the simplicity, let us take into account the following initial condition
\begin{equation}\label{Initial cond 3 spins}
\begin{aligned}
\rho(0)=&\ket{+}\bra{+}_1 \otimes [p_1\ket{++}\bra{++}+ \\
&+p_2\ket{+-}\bra{+-}+p_3\ket{-+}\bra{-+}+p_4\ket{--}\bra{--}].
\end{aligned}
\end{equation}
At the light of the previous discussion, considering a $\pi$-pulse, it is easy to verify that we may write the state at time $t$ as
\begin{equation}
\begin{aligned}
&\rho(t) \approx p_1\ket{-}\bra{-}_1\otimes \ket{--}\bra{--}+ \\
& +\ket{+}\bra{+}_1\otimes[p_2\ket{+-}\bra{+-}+p_3\ket{-+}\bra{-+}+p_4\ket{--}\bra{--}].
\end{aligned}
\end{equation}
Thus, a measurement act on the first spin projecting it in the state $\ket{-}_1$, has the effect to project all other spins too into their down-states.
Therefore, through an ancilla qubit and the specialized interaction model under scrutiny leading to a selective coupling, we may produce what we may call a \textit{selective-interaction-based cooling effect} of the spin system.
It is easy to understand that an analogue result may be obtained also for a greater odd number of spins. 

It is important to stress that the previous procedure and result are not valid in case of an even number of spins. This is due to the fact that, as we can see in Eq.~\eqref{HNPariSpinDiagonalizzata}, in each subdynamics we have an effective transverse magnetic field only along the $x$-direction and then the Rabi scenario with the related dynamics cannot be reproduced. However, additional appropriate conditions help to circumvent the $N$-parity constraint giving rise even in this case to a similar result of selective-interaction-based cooling effect. Let us consider, for simplicity, only the coupling in the $x$-direction, that is the following further simplification of Eq. \eqref{HNspinMessina}: $H(t)=\sum_{k=1}^N \hbar \omega_k\hat{\sigma}_k^z + \gamma_x(t) \prod_k \hat{\sigma}_k^x$. It is possible to see that if $\omega_1$ is sufficiently greater than all the other $\omega_k$ we may use the RWA \cite{Sakurai} in each subspace and then, in this instance, we restore the presence of Rabi oscillations in each subspace.
Matching the oscillation frequency of the coupling constant $\gamma_x(t)$ with the characteristic frequency of the subspace involving the states $\ket{+}^{\otimes N}$ and $\ket{-}^{\otimes N}$, namely $\nu=\sum_{k=1}^N\omega_k$, we obtain complete oscillations only in such a subspace.
The other sub-dynamics, instead, will be characterized by a frozen dynamics, provided that $\omega_k \gg \gamma, \forall k$. Therefore, if the system starts from the analogous state written in Eq. \eqref{Initial cond 3 spins}, we achieve also for an even number of spins the `cooling effect' thanks to the possibility to select a specific subspace in which the $N$-spin dynamics takes place. As a last remark it is worth to note that in this last case the result is based on the RWA, while in the different scenario for an odd number of spins, the result previously exposed is exact. It is interesting to note that this aspect can be seen also as an $N$-parity-dependent physical response of the system.

\section{Conclusions}

In this paper we have exactly solved a time-dependent model of $N$ spin-1/2 systems comprising highly non-local interactions.
Firstly, we have shown that, thanks to non-local $N$-order interaction terms, it is possible to reverberate to all the spins in the system the dynamical effects generated in one of the $N$ spins (ancilla qubit) by the application of a time-dependent magnetic field.
This allows us to generate easily GHZ sates or a contemporary perfect inversion of all the spins.
Secondly, we proposed a protocol through which we may generate a cooling effect of the whole spin system based on what we called selective interaction.
The latter consists in the possibility to select a specific dynamically invariant subspace for a non-trivial dynamics of the $N$-spin system, by appropriately engineering the time-dependence of the coupling parameters.

The key to get such physical results lies on the possibility to solve exactly the dynamics of the $N$-spin system by reducing the problem into a set of independent dynamical problems of single spin-1/2's.
As a final remark and outlook, it is worth noticing that this fact, identifiable as a result itself, makes possible the study of the dynamics of the system also when we consider random fluctuating components of the magnetic fields, as analogously done in \cite{GBNM}.
This would permit to analyse possible effects on the dynamics of the $N$ spins stemming from the coupling with an environment and to consider, then, situations closer to the experimental ones.

\section{Acknowledgements}
E.S. and L.L. acknowledge support from Ram\'on y Cajal Grant RYC-2012-11391, MINECO/FEDER FIS2015-69983-P, Basque Government IT986-16; R.G. acknowledges support from research funds difc 3100050001d08+, University of Palermo.

\appendix

\section{Transformation Procedure}\label{Transf Prod}

Let us consider the following $N$-spin model
\begin{equation}\label{HNspinMessinaSM}
H=\sum_{k=1}^{N}\hbar\omega_{k}\hat{\sigma}_{k}^{z}+\gamma_{x}\prod_{k=1}^{N}\hat{\sigma}_{k}^{x}+\gamma_{y}\prod_{k=1}^{N}\hat{\sigma}_{k}^{y}+\gamma_{z}\prod_{k=1}^{N}\hat{\sigma}_{k}^{z},
\end{equation}
describing $N$ distinguishable spins subjected to, in general, different magnetic fields and interacting between them only through $N$-wise interaction terms, that is each interaction term involves all the $N$-spins at the same time.
$\hat{\sigma}^x$, $\hat{\sigma}^y$ and $\hat{\sigma}^z$ are the standard Pauli matrices.

This model may be exactly diagonalized by a process consisting in a chain of unitary transformations.
To this end it is useful to start by considering the easiest case of two interacting spin 1/2's.
In this instance the Hamiltonian reads
\begin{equation}\label{H2SpinMessinaSM}
H_2= 
\hbar\omega_{1}\hat{\sigma}_{1}^{z}+\hbar\omega_{2}\hat{\sigma}_{2}^{z}+
\gamma_{x}\hat{\sigma}_{1}^{x}\hat{\sigma}_{2}^{x}+\gamma_{y}\hat{\sigma}_{1}^{y}\hat{\sigma}_{2}^{y}+\gamma_{z}\hat{\sigma}_{1}^{z}\hat{\sigma}_{2}^{z}
\end{equation}
and it is possible to verify that $[H_2,\hat{\sigma}_{1}^{z}\hat{\sigma}_{2}^{z}]=0$.
Transforming $H_2$ through the following unitary and hermitian operator ($\mathbb{1}$ is the identity operator in the four dimensional Hilbert subspace)
\begin{equation}\label{OperatoreUH2spinMessinaSM}
\mathbb{U}_{12}=\dfrac{1}{2}\left[\mathbb{1}+\hat{\sigma}_{1}^{z}+\hat{\sigma}_{2}^{x}-\hat{\sigma}_{1}^{z}\hat{\sigma}_{2}^{x}\right],
\end{equation}
we get
\begin{equation}\label{Htilde2SpinMessinaTransfSM}
\mathbb{U}_{12}^{\dagger}H_2\mathbb{U}_{12}=\tilde{H}_{2}=\hbar\left(\omega_{1}+\omega_{2}\hat{\sigma}_{2}^{z}\right)\hat{\sigma}_{1}^{z}+
\gamma_{x}\hat{\sigma}_{1}^{x}-\gamma_{y}\hat{\sigma}_{2}^{z}\hat{\sigma}_{1}^{x}+\gamma_{z}\sigma_{2}^{z}.
\end{equation}
It is easy to see that $\hat{\sigma}_{2}^{z}$ is constant of motion for $\tilde{H}$ and thus it may be treated as a parameter ($=\pm 1$), rewriting
\begin{equation}\label{Htilde2SpinMessinaTransfParamSM}
\tilde{H}_{\sigma_{2}^{z}}=\hbar\left(\omega_{1}+\omega_{2}\sigma_{2}^{z}\right)\hat{\sigma}_{1}^{z}+
\left(\gamma_{x}-\gamma_{y}\sigma_{2}^{z}\right)\hat{\sigma}_{1}^{x}+\gamma_{z}\sigma_{2}^{z}.
\end{equation}
This means that we have got two Hamiltonians of single spin-1/2, each one related to one of the two eigenvalues of $\hat{\sigma}_{2}^{z}$, $\pm 1$.
So, in this manner, we have reduced the two-interacting-spin problem into two independent single-spin-1/2 problems, easier to be solved.
Furthermore, it is worth to underline that each single-spin-1/2 Hamiltonian governs the dynamics of our two-spin system in one of the two dynamically invariant Hilbert subspace related to the two eigenvalue of $\hat{\sigma}_{2}^{z}$.

If we now consider the case of three spins, the Hamiltonian \eqref{HNspinMessinaSM} reads
\begin{equation}\label{H3spinMessinaSM}
\begin{aligned}
H_3 = 
&\hbar\omega_{1}\hat{\sigma}_{1}^{z}+\hbar\omega_{2}\hat{\sigma}_{2}^{z}+\hbar\omega_{3}\hat{\sigma}_{3}^{z}+ \\
&\gamma_{x}\hat{\sigma}_{1}^{x}\hat{\sigma}_{2}^{x}\hat{\sigma}_{3}^{x}+
\gamma_{y}\hat{\sigma}_{1}^{y}\hat{\sigma}_{2}^{y}\hat{\sigma}_{3}^{y}+
\gamma_{z}\hat{\sigma}_{1}^{z}\hat{\sigma}_{2}^{z}\hat{\sigma}_{3}^{z}.
\end{aligned}
\end{equation}
Now, it is possible to convince oneself that $\hat{\sigma}_{2}^{z}\hat{\sigma}_{3}^{z}$ is constant of motion and then if we apply the procedure previously used to the two spins 2 and 3 in $H_3$, we get the following new Hamiltonian
\begin{equation}\label{Htilde3spinMessinaSM}
\begin{aligned}
\mathbb{U}_{23}^{\dagger}H_3\mathbb{U}_{23}=&\tilde{H}_{3}=\hbar\omega_{1}\hat{\sigma}_{1}^{z}+
\hbar\left(\omega_{2}+\omega_{3}\sigma_{3}^{z}\right)\hat{\sigma}_{2}^{z}+ \\
&\gamma_{x}\hat{\sigma}_{1}^{x}\hat{\sigma}_{2}^{x}-\gamma_{y}\sigma_{3}^{z}\hat{\sigma}_{1}^{y}\hat{\sigma}_{2}^{x}+\gamma_{z}\sigma_{3}^{z}\hat{\sigma}_{1}^{z},
\end{aligned}
\end{equation}
where $\sigma_{3}^{z}$ (integral of motion) appears as parameter and so we have two different Hamiltonians of two interacting spin 1/2's.
This time the unitary and hermitian operator accomplishing the transformation is
\begin{equation}\label{OperatoreTrasfUH3spinMessinaSM}
\mathbb{U}_{23}=\dfrac{1}{2}\left[\mathbb{1}+\hat{\sigma}_{2}^{z}+\hat{\sigma}_{3}^{x}-\hat{\sigma}_{2}^{z}\hat{\sigma}_{3}^{x}\right],
\end{equation}
in accordance with the form of $\mathbb{U}_{12}$.
It is immediate, at this point, to understand that we may apply one more time the same procedure for $\tilde{H}_{3}$, using the operator written in Eq. \eqref{OperatoreUH2spinMessinaSM} since $\hat{\sigma}_{1}^{z}\hat{\sigma}_{2}^{z}$ is constant of motion for $\tilde{H}_3$.
Thus, we get
\begin{equation}\label{Hbitilde3spinMessinaSM}
\begin{aligned}
&\mathbb{U^{\dagger}}_{12}\tilde{H}_{3}\mathbb{U}_{12}=\mathbb{U^{\dagger}}_{123}H_{3}\mathbb{U}_{123}=\tilde{\tilde{H}}_{3}= \\
&\hbar\left(\omega_{1}+\omega_{2}\sigma_{2}^{z}+\omega_{3}\sigma_{2}^{z}\sigma_{3}^{z}\right)\hat{\sigma}_{1}^{z}+
\gamma_{x}\hat{\sigma}_{1}^{x}-\gamma_{y}\sigma_{3}^{z}\hat{\sigma}_{1}^{y}+\gamma_{z}\sigma_{3}^{z}\hat{\sigma}_{1}^{z},
\end{aligned}
\end{equation}
where we put $\mathbb{U}_{123}=\mathbb{U}_{23}\mathbb{U}_{12}$.
In this case we have two parameters, $\sigma_{2}^{z}$ and $\sigma_{3}^{z}$, and so we have four Hamiltonians of single spin-1/2 governing the dynamics of the three spin system in each of the four dynamically invariant subspaces related to the four pairs of the eigenvalues of the two constant of motion $\hat{\sigma}_{1}^{z}\hat{\sigma}_{2}^{z}$ and $\hat{\sigma}_{2}^{z}\hat{\sigma}_{3}^{z}$.
Therefore, also in this case, we have reduced the initial dynamical problem of three interacting spins into independent problems of a single spin-1/2.

Basing on this last result we understand that, for the case of $N$ spins, if we apply the procedure previously exposed for three spins, to the last three spins, we obtain a new Hamiltonian characterized by the same structure of the original one with the parameters redefined and depending only on the first $N-2$ spins (the last two spins appear as parameter).
One can imagine to iterate this procedure for each spin-triplet until the Hamiltonian is completely reduced to that of a single spin 1/2.
More precisely, it means that if we had, e.g., ten spins we could consider firstly the spin-triplet (8 9 10) and diagonalize the Hamiltonian with respect to these three spins, obtaining a new Hamiltonian depending on the dynamical variables of the spin 8 and those of the other spins not involved in the transformation; the spins 9 and 10 would appear only through $\sigma_{9}^{z}$ and $\sigma_{10}^{z}$ having the role of parameters.
At this point we should proceed by considering the spin-triplets (6 7 8), (4 5 6) and so on, diagonalizing every time with respect to the spin-triplet under consideration until we get a final Hamiltonian depending only on one spin 1/2, actually the first spin for the example taken into account.
It is important to underline that in the case of odd number of spins, through this technique, we get directly a final Hamiltonian of a single spin-1/2, while for an even number of spin we get firstly a Hamiltonian of two spins which can be treated analogously to get the final one depending on just one spin.

It is appropriate to define and make clear what we intend for ``diagonalize with respect to a spin-triplet''.
Considering the generic spin-triplet ($i$, $j$, $k$) (with $i<j<k$), diagonalizing with respect the three spins $i$, $j$ and $k$ means to transform the Hamiltonian through the following operator
\begin{equation}
\mathbb{U}_{ijk}=
\dfrac{1}{4}
\left[\mathbb{1}+\hat{\sigma}_{j}^{z}+\hat{\sigma}_{k}^{x}-\hat{\sigma}_{j}^{z}\hat{\sigma}_{k}^{x}\right]
\left[\mathbb{1}+\hat{\sigma}_{i}^{z}+\hat{\sigma}_{j}^{x}-\hat{\sigma}_{i}^{z}\hat{\sigma}_{j}^{x}\right]
\end{equation}
acting only upon the dynamical variables of the three spins under consideration.
As shown and explained before, this transformation leaves the Hamiltonian dependent on the dynamical variables of the first spin of the triplet ($i$-th spin) and on those of all the other spins not affected by the transformation.
The spins $j$ and $k$ appears only with $\hat{\sigma}_{j}^{z}$ and $\hat{\sigma}_{k}^{z}$ which, being constant of motion, may be treated as parameters and substituted with their eigenvalues in the expression of the transformed Hamiltonian.

It is useful now to observe what are the effects on the Hamiltonian after a diagonalization with respect to a spin-triplet:
\begin{itemize}
\item
a -1 factor appears in the interaction term in $\gamma_y$;

\item
the $\sigma^{z}$ operator (parameter) of the last spin in the triplet appears in the interaction terms in $\gamma_y$ and $\gamma_z$;

\item
the Pauli spin operators ($\hat{\sigma}^{x}$, $\hat{\sigma}^{y}$ and $\hat{\sigma}^{z}$) of the first spin of the triplet under consideration remain unchanged in each relative interaction term ($\gamma_x$, $\gamma_y$ and $\gamma_z$).
\end{itemize}
We observe also that, from Eqs. \eqref{Htilde2SpinMessinaTransfSM} and \eqref{Hbitilde3spinMessinaSM}, it is easy to conjecture the general form of the factor multiplying $\hat{\sigma}_{1}^{z}$ and depending on the $\omega_k$ parameters, namely
\begin{equation}
\omega_{1}+\sum_{k=2}^{N}\omega_{k}\prod_{k'=2}^{k}\sigma_{k'}^{z}.
\end{equation}

For, we are able, via an induction procedure, to write the argued form of the final single-spin-1/2 Hamiltonian.
\begin{widetext}
In the case of an odd number of spins it reads
\begin{equation}\label{HNdispariSpinDiagonalizzataSM}
\tilde{H}=\hbar\Biggl[\omega_{1}+\sum_{k=2}^{N}\omega_{k}\prod_{k'=2}^{k}\sigma_{k'}^{z}\Biggr]\hat{\sigma}_{1}^{z}+\gamma_{x}\hat{\sigma}_{1}^{x}+\Biggl[(-1)^{N-1\over 2}\gamma_{y}\prod_{k=1}^{(N-1)/2}\sigma_{2k+1}^{z}\Biggr]\hat{\sigma}_{1}^{y}+\Biggl[\gamma_{z}\prod_{k=1}^{(N-1)/2}\sigma_{2k+1}^{z}\Biggr]\hat{\sigma}_{1}^{z},
\end{equation}
whereas for an even number of spins we have
\begin{equation}\label{HNPariSpinDiagonalizzataSM}
\tilde{H}=\hbar\Biggl[\omega_{1}+\sum_{k=2}^{N}\omega_{k}\prod_{k'=2}^{k}\sigma_{k'}^{z}\Biggr]\hat{\sigma}_{1}^{z}+\gamma_{x}\hat{\sigma}_{1}^{x}+\Biggl[(-1)^{N \over 2}\gamma_{y}\prod_{k=1}^{N/2}\sigma_{2k}^{z}\Biggr]\hat{\sigma}_{1}^{x}+\gamma_{z}\prod_{k=1}^{N/2}\sigma_{2k}^{z}.
\end{equation}
\end{widetext}
It is of relevance to underline that ${(N-1)/2}$ and $N/2$, appearing respectively in Eq. \eqref{HNdispariSpinDiagonalizzataSM} and \eqref{HNPariSpinDiagonalizzataSM}, are the numbers of transformations to be applied to the original Hamiltonian in Eq. \eqref{HNspinMessinaSM} to get the final ones.
The total unitary operator accomplishing this chained transformations may be written as
\begin{equation}\label{OperatoreDiagonalizzanteHNspinSM}
\mathbb{U}
=\dfrac{1}{2^{N-1}}\prod_{k=0}^{N-2}\left[\mathbb{1}+\hat{\sigma}_{N-(k-1)}^{z}+\hat{\sigma}_{N-k}^{x}-\hat{\sigma}_{N-(k+1)}^{z}\hat{\sigma}_{N-k}^{x}\right].
\end{equation}

\section{Eigenvectors and Breaking Down of the Schr\"odinger Equation} \label{SR Eq}

To understand the eigenvectors structure, let us consider, for the sake of simplicity, the simplest case of two spin-1/2's.
By Eqs. \eqref{Htilde2SpinMessinaTransfSM} and \eqref{Htilde2SpinMessinaTransfParamSM}, it is easy to understand that we may write the eigenvectors of $\tilde{H}$ as follows
\begin{equation}\label{tensprodautostatiHtildeSM}
\ket{\tilde{\psi}_{ij}}=\ket{\phi_{ij}} \otimes \ket{\sigma_{2}^{z}=i}
\end{equation}
with $i=\pm1$, $j=1,2$, $\ket{\sigma_{2}^{z}=1}=(1,0)^T$ and $\ket{\sigma_{2}^{z}=-1}=(0,1)^T$.
In the previous expressions, $\ket{\phi_{1i}}$ ($\ket{\phi_{-1i}}$) are the two eigenvectors of $\tilde{H}_{+1}$ ($\tilde{H}_{-1}$).
Finally, the eigenvectors of $H$ are easily derived by the relation
\begin{equation}
\mathbb{U}\tilde{\ket{\psi_{ij}}}=\ket{\psi_{i}}.
\end{equation}

If the Hamiltonian $H$ is time-dependent, we have to study the time-dependent Schr\"odinger equation, namely
\begin{equation}
i\hbar \ket{\dot{\psi}(t)}=H(t) \ket{\psi(t)}.
\end{equation}
Since $\frac{\partial}{\partial t} \mathbb{U}=0$, it is easy to verify that we may write
\begin{equation}
i\hbar \ket{\dot{\tilde{\psi}}(t)}=\tilde{H}(t) \ket{\tilde{\psi}(t)}.
\end{equation}
By writing a general initial condition as follows
\begin{equation}
\ket{\tilde{\psi}(0)}=
\begin{pmatrix}
a \\ b\\ c\\ d
\end{pmatrix}=
\begin{pmatrix}
a \\ c
\end{pmatrix} \otimes
\begin{pmatrix}
1 \\ 0
\end{pmatrix}+
\begin{pmatrix}
b \\ d
\end{pmatrix} \otimes
\begin{pmatrix}
0 \\ 1
\end{pmatrix},
\end{equation}
since $[\tilde{H}(t),\sigma_{2}^{z}]=0$, we may write the evolved state at time $t$ as
\begin{equation}
\begin{aligned}
\ket{\tilde{\psi}(t)}&=
\begin{pmatrix}
a(t) \\ b(t) \\ c(t) \\ d(t)
\end{pmatrix}=
\begin{pmatrix}
a(t) \\ c(t)
\end{pmatrix} \otimes
\begin{pmatrix}
1 \\ 0
\end{pmatrix}+
\begin{pmatrix}
b(t) \\ d(t)
\end{pmatrix} \otimes
\begin{pmatrix}
0 \\ 1
\end{pmatrix}=\\
&=\ket{\tilde{\phi_{1}}}_{1} \otimes \ket{\sigma_{2}^{z}=1}_{2} + \ket{\tilde{\phi}_{-1}}_{1} \otimes \ket{\sigma_{2}^{z}=-1}_{2}
\end{aligned}
\end{equation}
where $\ket{\tilde{\phi}_{\pm 1}}_{1}$ satisfy the following dynamical problems
\begin{equation}
i\hbar \ket{\dot{\tilde{\phi}}_{\pm 1}(t)}= \tilde{H}_{\pm 1}(t) \ket{\tilde{\phi}_{\pm 1}(t)}
\end{equation}
being nothing but two independent single spin-1/2 time-dependent Schr\"odinger equations.

\end{document}